\documentclass[12pt]{article}

\usepackage[margin=2.5cm]{geometry}
\usepackage[T1]{fontenc}
\usepackage[utf8]{inputenc}
\usepackage{amsmath}
\usepackage{amssymb}
\usepackage{amsthm}
\usepackage{bm}
\usepackage{graphicx}
\usepackage{booktabs}
\usepackage{placeins}
\usepackage{xcolor}
\usepackage{url}
\usepackage{hyperref}

\makeatletter
\renewcommand{\fnum@figure}{\textbf{Figure \thefigure}}
\renewcommand{\fnum@table}{\textbf{Table \thetable}}
\makeatother

\usepackage{caption}
\usepackage{natbib}
\definecolor{linkblue}{HTML}{023C85}
\hypersetup{
  colorlinks=true,
  linkcolor=linkblue,
  citecolor=linkblue,
  urlcolor=linkblue
}

\newcommand{\R}{\mathbb{R}}
\newcommand{\E}{\mathbb{E}}
\newcommand{\e}{{\rm e}}
\newcommand{\dd}{{\rm d}}

\newtheorem{proposition}{Proposition}
\newtheorem{corollary}{Corollary}
\newtheorem{lemma}{Lemma}
\theoremstyle{definition}
\newtheorem{remark}{Remark}

\graphicspath{{figures/}}
\title{An Exact Poisson Mixture Representation of Siler Remaining
Life Expectancy}
\author{Silvio C. Patricio\thanks{Email: \texttt{silca [at] sam.sdu.dk}. ORCID: \texttt{0000-0002-1025-5571}}\\[3pt]
\small Interdisciplinary Center on Population Dynamics\\
\small University of Southern Denmark}
\date{}

\begin{document}
\maketitle

\vspace{-10mm}

\begin{abstract} \noindent
The Siler model describes mortality from birth to old age as the sum of a declining immaturity hazard, a constant background hazard, and an increasing senescence hazard. Its remaining life expectancy has no known closed form, and numerical evaluation alone does not reveal how the immaturity hazard enters it. We show that the remaining-lifetime distribution at age $x$ is Gompertz--Makeham with a Poisson-distributed Makeham term. Successive Makeham terms differ by the rate at which immaturity mortality declines, and the Poisson mean is the cumulative immaturity hazard remaining after age $x$. In a latent-clock construction, individuals carry a Poisson number of constant-rate clocks, and selection removes those carrying more clocks first. Integrating the mixture yields an exact series for remaining life expectancy with a closed-form truncation bound. The life expectancy lost to immaturity mortality is close to the probability of carrying a clock times life expectancy with that hazard removed, with a closed-form error bound. In annual fits to Swedish female mortality, this loss at birth fell from 4.3 years in 1900 to less than 0.01 years in 2020, a decline equal to 15\% of the gain in life expectancy at birth, and the approximation was always within 0.06 years.
\end{abstract}


\section{Introduction}

\citet{Siler1979} wrote mortality as the sum of three competing risks: an immaturity hazard that falls exponentially from birth, a constant background hazard, and a senescence hazard that rises exponentially with age. With five parameters, the model describes the whole life course and has been fitted to human and animal populations alike \citep{Siler1983,LemaitreEtAl2020}. Remaining life expectancy summarizes such a fit. Under the competing-risk interpretation, it also determines the probability of eventually dying from the background risk, which equals the background hazard times remaining life expectancy \citep{PatricioMissov2024}. The Siler survival function is explicit, but its integral has no known closed form.

Without the immaturity hazard, the Siler model reduces to the Gompertz--Makeham model \citep{Makeham1860}, whose remaining life expectancy is an incomplete gamma function of the senescence hazard at the starting age. The immaturity hazard prevents this reduction. It declines with age while the senescence hazard grows, so the two exponential terms do not combine into a single incomplete gamma function. Numerical quadrature evaluates remaining life expectancy readily, but it does not show how the declining early-life hazard shapes the remaining-lifetime distribution, why that hazard declines, or how much life expectancy it costs.

We address these questions with a latent-clock representation of the immaturity hazard. It treats each individual as if they carried, from birth, a random number of hidden vulnerabilities, or latent clocks. Each clock can cause death at the same constant rate. An individual's immaturity hazard then does not decline with age: it equals the number of clocks carried times their common rate. Conditional on that number, the remaining lifetime is Gompertz--Makeham, with the combined clock rate added to the background hazard. The number of clocks is Poisson distributed, so the Siler remaining-lifetime distribution is a Poisson mixture of Gompertz--Makeham distributions. Its mixing mean is the cumulative immaturity hazard that remains at the starting age, which falls with age because individuals carrying more clocks die first.

This identity reduces calculations under the Siler model to Gompertz--Makeham ones. We integrate it to obtain an exact series for remaining life expectancy and bound its truncation error in years. We then use the same bound to relate the life expectancy lost to immaturity mortality to the probability of carrying at least one clock. We illustrate the results with annual Siler fits to Swedish female mortality from 1900 to 2020.

\section{The Siler model}\label{sec:siler}

Let $T$ denote age at death. The Siler model specifies the hazard
\[
\mu(t)=\mu_1(t)+a_2+\mu_3(t),
\qquad
\mu_1(t)=a_1\e^{-b_1t},
\qquad
\mu_3(t)=a_3\e^{b_3t},
\]
with $a_1,a_2\geq0$ and $a_3,b_1,b_3>0$. The immaturity hazard $\mu_1$ declines at rate $b_1$, $a_2$ is the background hazard, and the senescence hazard $\mu_3$ increases at rate $b_3$. For a life aged $x$, the remaining lifetime $T_x=T-x\mid T>x$ has survival function
\begin{equation}
{}_tp_x
=\exp\left\{-\int_x^{x+t}\mu(u)\,\dd u\right\}
=\exp\left\{
-A_x\left(1-\e^{-b_1t}\right)-a_2t-G_x\left(\e^{b_3t}-1\right)
\right\},
\label{eq:conditional-survival}
\end{equation}
where
\[
A_x=\int_x^\infty\mu_1(u)\,\dd u=\frac{a_1\e^{-b_1x}}{b_1},
\qquad
G_x=\frac{\mu_3(x)}{b_3}=\frac{a_3\e^{b_3x}}{b_3}.
\]
Here $A_x$ is the remaining cumulative immaturity hazard after age $x$, and $\e^{-A_x}$ is the probability of surviving it. The remaining life expectancy at age $x$ is
\begin{equation}
e_x=\E(T_x)=\int_0^\infty {}_tp_x\,\dd t.
\label{eq:rle-definition}
\end{equation}
Conditioning on survival to age $x$ replaces $a_1$ and $a_3$ by $\mu_1(x)$ and $\mu_3(x)$ and leaves $b_1$, $a_2$, and $b_3$ unchanged. The remaining lifetime at every age therefore follows a Siler law, and the results below hold for every $x$.

\section{The Siler model as a Poisson mixture}\label{sec:mixture}

A hazard that declines in the population can arise from individual hazards that do not decline, provided individuals differ in their level of risk. We use this idea to write the Siler immaturity hazard as heterogeneity in the number of latent clocks that individuals carry.

Let $N_0\sim\operatorname{Poisson}(A_0)$ be the number of latent clocks carried at birth. Conditional on $N_0$, the clocks ring, that is, cause death, at independent exponential times $C_1,\ldots,C_{N_0}$ with rate $b_1$, and they are independent of the background and senescence failure times. The immaturity failure time is the first ring, $C=\min_j C_j$, with $C=\infty$ when $N_0=0$. Its survival function is
\[
\Pr(C>t)=\E(\e^{-b_1N_0t})
=\exp\{-A_0(1-\e^{-b_1t})\},
\]
which is the survival function of the Siler immaturity hazard. Survival to age $x$ carries information about the clock count. By Bayes' rule, with $A_x=A_0\e^{-b_1x}$, it exponentially tilts the Poisson probabilities:
\[
\begin{aligned}
\Pr(N_0=n\mid C>x)
&\propto \Pr(N_0=n)\Pr(C>x\mid N_0=n) \\
&=\e^{-A_0}\frac{(A_0\e^{-b_1x})^n}{n!}
\propto\frac{(A_x)^n}{n!}.
\end{aligned}
\]
Survival from the background and senescence risks carries no information about $N_0$, so the same law holds among those who survive all three risks to age $x$. Let $K_x$ denote the number of clocks carried by a survivor at age $x$; then $K_x\sim\operatorname{Poisson}(A_x)$. A survivor carrying $K_x$ clocks has immaturity hazard $b_1K_x$, and the population immaturity hazard is its average among survivors:
\[
\mu_1(x)=b_1\E(K_x)=b_1A_x=a_1\e^{-b_1x}.
\]
This identity gives the declining hazard a selection interpretation. Each clock has constant rate $b_1$, but individuals carrying more clocks die sooner, so the mean clock count among survivors, and with it the population hazard, falls exponentially at rate $b_1$. \citet{VaupelYashin1985} show that a mixture of constant individual hazards produces a declining population hazard in this way. This heterogeneity differs from the gamma frailty of \citet{VaupelMantonStallard1979}, which multiplies the individual hazard and slows the rise of senescent mortality. Here the clocks add to the immaturity hazard alone, and their number follows a Poisson rather than a gamma distribution. Not every declining hazard admits a clock representation of this kind. Appendix~\ref{app:constant-rate} shows that an immaturity hazard can be generated by clocks with constant, possibly different, rates if and only if it is completely monotone, that is, a mixture of exponentially declining hazards. The Siler hazard is the case in which all clocks share one rate.

In evolutionary demography, the decline of mortality during development is called ontogenescence, and \citet{levitis2011before} lists heterogeneous frailty among its explanations. Under this reading, a clock is a persistent early-life vulnerability that raises the individual hazard by $b_1$ until death. The construction is an exact representation of the population survival function rather than an identified biological mechanism, because aggregate survival cannot distinguish it from mechanisms in which individual risk itself declines with development.

\begin{proposition}[Distributional Poisson mixture]\label{prop:distribution-mixture}
For every $x,t\geq0$,
\begin{equation}
{}_tp_x
=\E\!\left[
\exp\left\{-(a_2+b_1K_x)t-G_x(\e^{b_3t}-1)\right\}
\right],
\qquad K_x\sim\operatorname{Poisson}(A_x).
\label{eq:distribution-mixture}
\end{equation}
Thus the Siler remaining lifetime at age $x$ is a Poisson mixture of Gompertz--Makeham remaining lifetimes with Makeham term $a_2+b_1K_x$.
\end{proposition}

\begin{proof}
By memorylessness, conditional on $K_x=n$, the residual clock times are independent exponentials with rate $b_1$, so their joint survival over the next $t$ years is $\e^{-nb_1t}$. By the assumed independence of the competing risks, multiplying this factor by the background and senescence survival functions gives the integrand in Equation~\eqref{eq:distribution-mixture}; averaging over $K_x$ gives ${}_tp_x$.
\end{proof}

Proposition~\ref{prop:distribution-mixture} applies immediately to functionals that are linear in the remaining-lifetime distribution, which include remaining life expectancy, moments, the continuous life annuity, and the Laplace transform. Quantiles and other nonlinear functionals are determined by the mixture distribution but are not, in general, averages of their component values.

This representation complements the competing-risk mixture in \citet{PatricioMissov2024}. Their mixture conditions on the cause that produces death, whereas Proposition~\ref{prop:distribution-mixture} mixes over the number of clocks carried at age $x$. The constant-shift argument does not depend on the Gompertz form of the senescence hazard, and Appendix~\ref{app:general-adult} gives the corresponding result for a general adult hazard and its gamma--Gompertz specialization.

To obtain the remaining life expectancy explicitly, we write the relevant integrals in terms of the generalized exponential, or integro-exponential, function
\[
E_\nu(z)=\int_1^\infty u^{-\nu}\e^{-zu}\,\dd u
=z^{\nu-1}\Gamma(1-\nu,z),
\qquad z>0,\ \nu\in\R,
\]
where $\Gamma(s,z)=\int_z^\infty v^{s-1}\e^{-v}\,\dd v$ is the upper incomplete gamma function \citep{Milgram1985,OlverEtAl2010}. Writing $e_x^{\mathrm{GM}}(c)$ for remaining life expectancy under the Gompertz--Makeham hazard $c+a_3\e^{b_3t}$, the substitution $u=\e^{b_3t}$ gives
\begin{equation}
e_x^{\mathrm{GM}}(c)
=\frac{\e^{G_x}}{b_3}E_{1+c/b_3}(G_x)
=\frac{\e^{G_x}}{b_3}\,(G_x)^{c/b_3}\,\Gamma\!\left(-\frac{c}{b_3},G_x\right),
\label{eq:gm}
\end{equation}
the known expression for Gompertz--Makeham remaining life expectancy \citep{CastellaresEtAl2020}.

\begin{corollary}[Siler remaining life expectancy]\label{cor:mixture}
Let $w_n(x)=\Pr(K_x=n)=\e^{-A_x}(A_x)^n/n!$, and $Q_n(x)=e_x^{\mathrm{GM}}(a_2+nb_1)$. Then
\begin{equation}
e_x=\E\!\left[e_x^{\mathrm{GM}}(a_2+b_1K_x)\right]
=\sum_{n=0}^\infty w_n(x)\,Q_n(x),
\label{eq:gm-mixture}
\end{equation}
and
\begin{equation}
e_x=\frac{\e^{G_x-A_x}}{b_3}
\sum_{n=0}^\infty\frac{(A_x)^n}{n!}\,E_{1+(a_2+nb_1)/b_3}(G_x).
\label{eq:main-series}
\end{equation}
\end{corollary}

\begin{proof}
Integrating Equation~\eqref{eq:distribution-mixture} and applying Tonelli's theorem gives the expectation in Equation~\eqref{eq:gm-mixture}; expanding that expectation gives the sum. Substituting Equation~\eqref{eq:gm} into the sum gives Equation~\eqref{eq:main-series}.
\end{proof}

When $a_1=0$, $A_x=0$ and only the component $n=0$ remains. In general, the leading term
\[
e_x^{(0)}=\e^{-A_x}\,e_x^{\mathrm{GM}}(a_2)
\]
is the Gompertz--Makeham life expectancy with Makeham term $a_2$, weighted by the probability of carrying no clock. As $x$ increases, $A_x$ decays at rate $b_1$, so the mixing law concentrates on $K_x=0$ and $e_x$ approaches the Gompertz--Makeham value.

The clock representation also separates carrying an immaturity risk from dying from it. Let $\pi_1(x)$ denote the probability that a life aged $x$ eventually dies from the immaturity risk. Conditional on $K_x=n$, the immaturity hazard is the constant $nb_1$, so this probability is $nb_1Q_n(x)$. Averaging over $K_x$ gives
\[
\pi_1(x)=b_1\E\!\left[K_xQ_{K_x}(x)\right]
=b_1A_x\E\!\left[Q_{K_x+1}(x)\right].
\]
The second equality uses $n w_n(x)=A_x w_{n-1}(x)$.

By contrast, $\Pr(K_x\geq1)=1-\e^{-A_x}$ is the probability of carrying at least one clock, not the probability that a clock rings before the competing risks. The background probability is $a_2e_x$, as in \citet{PatricioMissov2024}, and the senescence probability is the remainder.

\begin{remark}[Laplace transform]\label{rem:laplace}
Discounting at force of interest $\delta\geq0$ multiplies ${}_tp_x$ by $\e^{-\delta t}$, which by Equation~\eqref{eq:conditional-survival} gives the survival function of a Siler model with background hazard $a_2+\delta$ and the same $A_x$ and $G_x$. Corollary~\ref{cor:mixture} therefore gives the continuous life annuity
\[
\int_0^\infty\e^{-\delta t}\,{}_tp_x\,\dd t
=\E\!\left[e_x^{\mathrm{GM}}(a_2+\delta+b_1K_x)\right],
\]
as a Poisson mixture of the Gompertz--Makeham annuities of \citet{CastellaresPatricioLemonte2022}. Because $\E(\e^{-\delta T_x})=1-\delta\int_0^\infty\e^{-\delta t}\,{}_tp_x\,\dd t$, the Laplace transform of the remaining lifetime is the same mixture of Gompertz--Makeham Laplace transforms.
\end{remark}

\section{Truncation error}\label{sec:truncation}

Each mixture component is a life expectancy under a total hazard of at least $a_2+nb_1$, so it has a simple bound.

\begin{lemma}\label{lem:qn}
The sequence $Q_n(x)$ is nonincreasing in $n$, and
\[
Q_n(x)\leq\frac{1}{a_2+nb_1},
\]
whenever $a_2+nb_1>0$.
\end{lemma}

\begin{proof}
By the definition of $Q_n(x)$,
\[
Q_n(x)=\int_0^\infty
\exp\left\{-(a_2+nb_1)t-G_x\left(\e^{b_3t}-1\right)\right\}\,\dd t.
\]
The integrand decreases in $n$ and is bounded by $\e^{-(a_2+nb_1)t}$.
\end{proof}

\begin{proposition}\label{prop:truncation}
For $m\geq0$, let $e_x^{(m)}=\sum_{n=0}^{m}w_n(x)Q_n(x)$. Then $e_x^{(m)}\uparrow e_x$ as $m\to\infty$, and
\begin{equation}
0\leq e_x-e_x^{(m)}\leq
\frac{\Pr(K_x>m)}{a_2+(m+1)b_1}.
\label{eq:truncation-bound}
\end{equation}
\end{proposition}

\begin{proof}
By Corollary~\ref{cor:mixture}, $e_x-e_x^{(m)}=\sum_{n>m}w_n(x)Q_n(x)$. All terms are nonnegative, so $e_x^{(m)}$ is nondecreasing in $m$ and converges to $e_x$. By Lemma~\ref{lem:qn}, $Q_n(x)\leq1/[a_2+(m+1)b_1]$ for every $n>m$. Hence
\[
0\leq e_x-e_x^{(m)}
\leq\frac{1}{a_2+(m+1)b_1}\sum_{n>m}w_n(x)
=\frac{\Pr(K_x>m)}{a_2+(m+1)b_1}.
\]
\end{proof}

For the leading term, Equation~\eqref{eq:truncation-bound} reads
\begin{equation}
0\leq e_x-\e^{-A_x}e_x^{\mathrm{GM}}(a_2)
\leq\frac{1-\e^{-A_x}}{a_2+b_1}
\leq\frac{A_x}{a_2+b_1}.
\label{eq:leading-term-bound}
\end{equation}
The leading term is therefore accurate when little cumulative immaturity hazard remains or when immaturity mortality declines rapidly. The same bounds hold for the integral in Remark~\ref{rem:laplace}, with $a_2$ replaced by $a_2+\delta$.

Removing the immaturity hazard turns the Siler model into the Gompertz--Makeham model. Thus $e_x^{\mathrm{GM}}(a_2)$ is the cause-deleted remaining life expectancy, and $\Delta_x=e_x^{\mathrm{GM}}(a_2)-e_x$ is the gain in remaining life expectancy from deleting immaturity mortality \citep{Keyfitz1977,BeltranSanchezEtAl2008}. Rearranging Equation~\eqref{eq:leading-term-bound} gives
\begin{equation}
\left(1-\e^{-A_x}\right)
\left[e_x^{\mathrm{GM}}(a_2)-\frac{1}{a_2+b_1}\right]
\leq\Delta_x\leq
\left(1-\e^{-A_x}\right)e_x^{\mathrm{GM}}(a_2).
\label{eq:loss-bounds}
\end{equation}
Here $1-\e^{-A_x}=\Pr(K_x\geq1)$ is the probability of carrying at least one clock. Dividing the remainder in Equation~\eqref{eq:leading-term-bound} by this probability gives $\E[Q_{K_x}(x)\mid K_x\geq1]$, the mean remaining lifetime under all three risks among clock carriers. Every carrier has background-plus-immaturity hazard at least $a_2+b_1$, so this conditional mean is at most $1/(a_2+b_1)$. Hence the mean gain from deleting immaturity mortality among clock carriers lies between $e_x^{\mathrm{GM}}(a_2)-1/(a_2+b_1)$ and $e_x^{\mathrm{GM}}(a_2)$.

\section{Application to Swedish female mortality}

We use female death counts $D_x$ and exposures $E_x$ for Sweden, 1900--2020, at single ages 0--109, from the Human Mortality Database \citep{HMD2026}. For each year, we assume $D_x\sim\operatorname{Poisson}(E_x\mu(x))$ independently across ages \citep{Brillinger1986} and estimate $\bm\theta=(a_1,b_1,a_2,a_3,b_3)$ by maximizing the log-likelihood
\[
\ell(\bm\theta)
=\sum_x\left\{D_x\log\mu(x)-E_x\mu(x)\right\}.
\]

\begin{figure}[!htb]
\centering
\includegraphics[width=\textwidth]{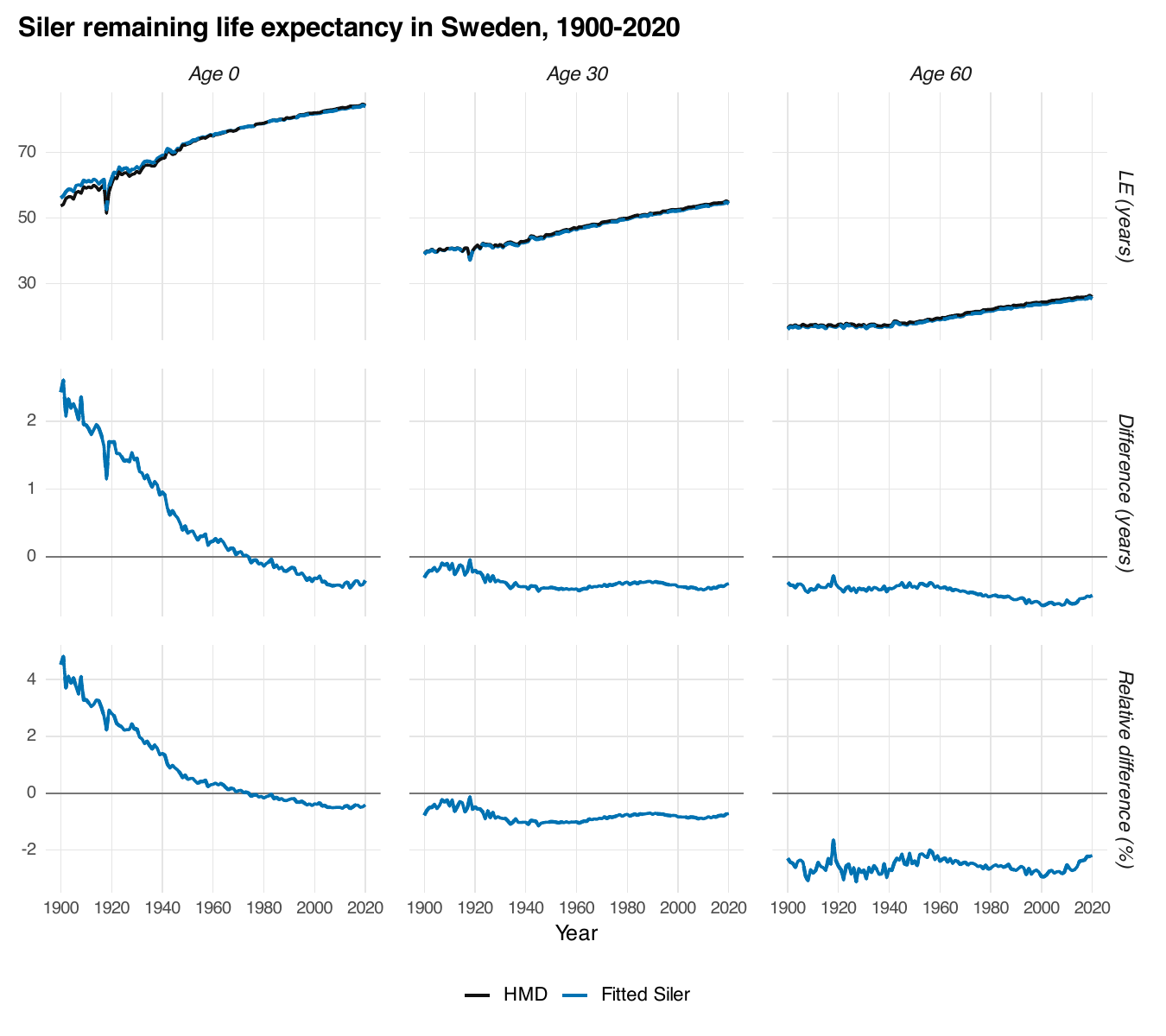}
\caption{Remaining life expectancy for Swedish females, 1900--2020, at ages 0, 30, and 60. Top row: HMD life-table values and fitted Siler values. Middle row: fitted minus HMD, in years. Bottom row: the same difference as a percentage of the HMD value.}
\label{fig:remaining-life}
\end{figure}

Figure~\ref{fig:remaining-life} compares the fitted Siler remaining life expectancy with HMD values at ages 0, 30, and 60. The fitted model overstates $e_0$ by up to 2.6 years before 1940 and understates $e_{60}$ by 1.7--3.1\% throughout. Part of the gap is definitional, since HMD life tables smooth death rates at old ages and use specific assumptions for the average time lived in the first year of life by infants who die. We use the fits to evaluate the series and the loss bounds rather than to assess the Siler model as a description of Swedish mortality.

\subsection{The loss to immaturity mortality nearly vanished}

\begin{figure}[!htb]
\centering
\includegraphics[width=\textwidth]{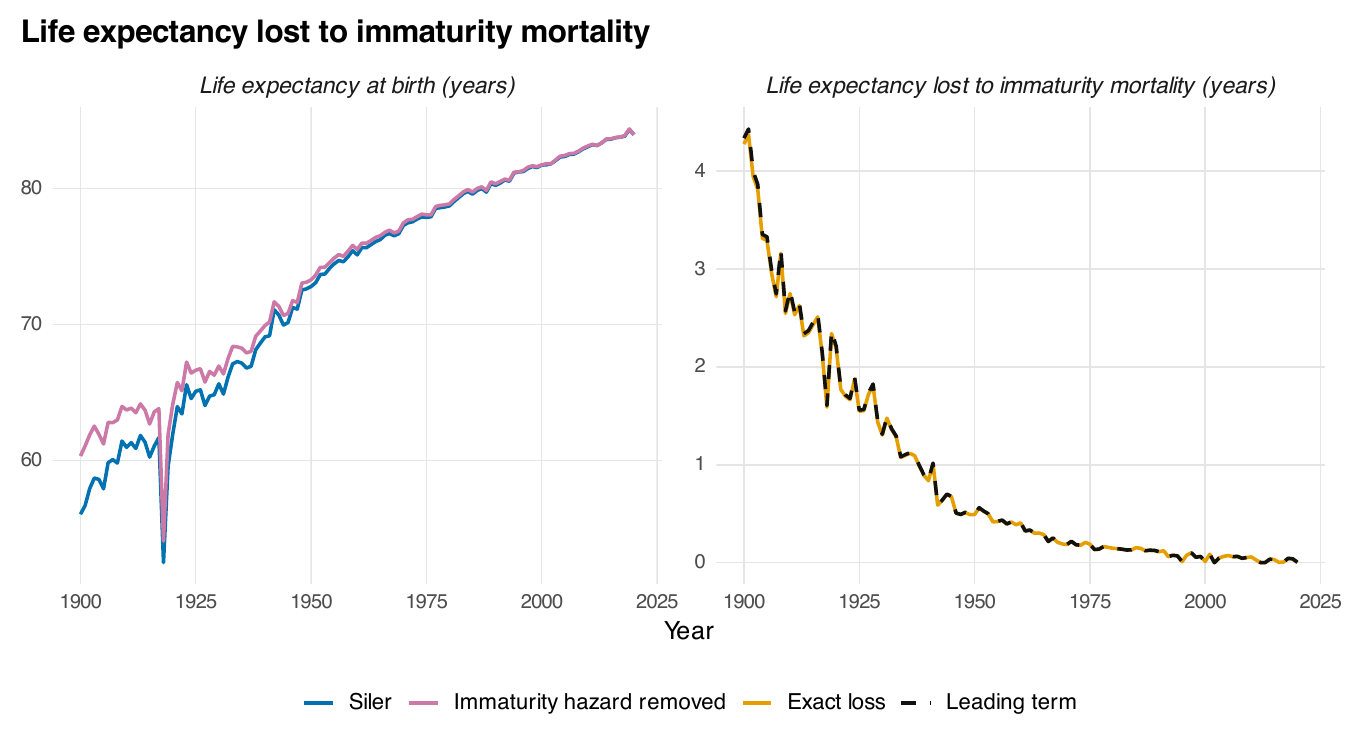}
\caption{Swedish females, 1900--2020. Left: fitted Siler life expectancy at birth $e_0$ and the Gompertz--Makeham life expectancy with the immaturity hazard removed, $e_0^{\mathrm{GM}}(a_2)$. Right: the loss $\Delta_0=e_0^{\mathrm{GM}}(a_2)-e_0$ and its leading-term value $(1-\e^{-A_0})e_0^{\mathrm{GM}}(a_2)$.}
\label{fig:immaturity-loss}
\end{figure}

The life expectancy lost to immaturity mortality fell from 4.28 years in 1900 to 0.49 years in 1950 and 0.004 years in 2020 (Figure~\ref{fig:immaturity-loss}). Over the same period, $e_0$ rose by 27.9 years and $e_0^{\mathrm{GM}}(a_2)$ by 23.6 years, so under this cause-deleted decomposition the decline of immaturity mortality accounts for 15\% of the gain. The leading-term approximation is close in every year: $\Delta_0$ lies between 98.6\% and 99.99\% of its upper bound $(1-\e^{-A_0})e_0^{\mathrm{GM}}(a_2)$ in Equation~\eqref{eq:loss-bounds} and differs from that bound by at most 0.060 years. The probability of carrying at least one clock, multiplied by life expectancy with the immaturity hazard removed, therefore approximates the loss at birth closely.

\subsection{A few terms of the series suffice}

We evaluate $E_\nu(z)$ through the incomplete gamma function in the R package \texttt{expint} \citep{Goulet2026}, falling back to numerical integration of
\[
\e^zE_\nu(z)=\int_0^\infty
\exp\{-z(\e^s-1)-(\nu-1)s\}\,\dd s
\]
when the routine returns a nonfinite value. We stop the series at the smallest $m$ for which the bound in Equation~\eqref{eq:truncation-bound} is below $10^{-12}$ years. For the years 1920, 1940, \ldots, 2020 and ages 0, 30, and 60, the series and adaptive quadrature of Equation~\eqref{eq:rle-definition} differ by at most $1.08\times10^{-12}$ years (Table~\ref{tab:numerical-verification}). These differences come from the numerical routines, not from truncation: at ages 30 and 60, the truncation bound is below $10^{-16}$ years.

\begin{figure}[!htb]
\centering
\includegraphics[width=\textwidth]{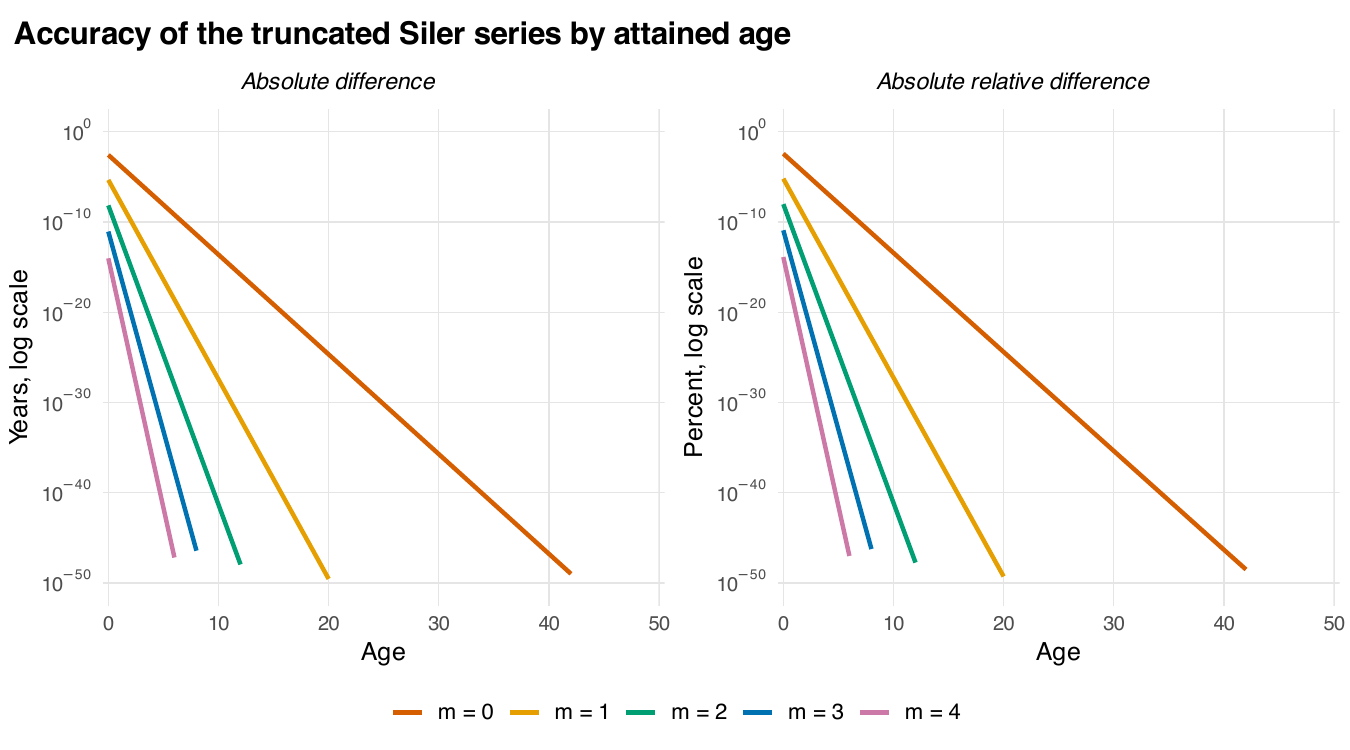}
\caption{Truncation error of the Siler series for Swedish females in 1950: $|e_x^{(m)}-e_x|$ in years (left) and as a percentage of $e_x$ (right), $m=0,\ldots,4$. The errors are evaluated by summing the nonnegative mixture terms beyond $m$. Logarithmic scales; values below $10^{-50}$ are omitted.}
\label{fig:series-convergence}
\end{figure}

Figure~\ref{fig:series-convergence} shows the error of the partial sums $e_x^{(m)}$, $m=0,\ldots,4$, at ages 0--50 in 1950, and Figure~\ref{fig:maximum-approximation-difference} in Appendix~\ref{app:numerical-verification} shows its maximum over all years at each age from 0 to 100. The natural logarithm of the error declines almost linearly with age, with slope close to $-(m+1)b_1$, because the omitted tail is dominated by a term proportional to $(A_x)^{m+1}$. On the base-10 scale of the figures, the slope is $-(m+1)b_1/\log 10$. Over all years and ages, the leading term alone is within $0.0601$ years ($0.107\%$) of $e_x$. Each further term lowers the maximum error by a factor of 50 to 85, to $1.12\times10^{-3}$, $1.85\times10^{-5}$, $2.60\times10^{-7}$, and $3.12\times10^{-9}$ years for $m=1,\ldots,4$.

\section{Discussion and conclusion}

The Siler remaining lifetime is Gompertz--Makeham with a Poisson-distributed Makeham term. This identity separates the two roles of the immaturity hazard. Its remaining cumulative hazard $A_x$ sets the expected clock count among survivors, while $b_1$ is both the rate of each clock and the increment in the Makeham term between successive components. Adaptive quadrature suffices for numerical evaluation, but the series adds a guaranteed error in years and isolates the immaturity contribution.

The same representation yields a simple demographic approximation. The life expectancy lost to immaturity mortality is close to the probability of carrying at least one clock multiplied by life expectancy with the immaturity hazard removed. Equation~\eqref{eq:loss-bounds} bounds the approximation for every parameter vector, because carriers survive on average for at most $1/(a_2+b_1)$ years under all three risks. In the Swedish fits, the approximation is within 0.06 years at every age and in every year. Under the cause-deleted decomposition, the decline in immaturity mortality contributed about 4.3 of the 27.9 years gained in life expectancy at birth between 1900 and 2020.

The construction also gives the declining limb of the Siler curve an exact frailty representation, which corresponds to one of the explanations for ontogenescence reviewed by \citet{levitis2011before}. The characterization in Appendix~\ref{app:constant-rate} gives a check for any fitted immaturity hazard: if the hazard is not completely monotone, its decline cannot arise from constant-rate clocks. For a stretched-exponential immaturity hazard with $0<p<1$, clock rates are random, so the persistent vulnerabilities may differ in lethality, and the proportional rate of decline of the immaturity hazard decreases with age.

Three caveats bound these results. First, the biological interpretation is not identifiable from aggregate mortality. The same population curve may arise when individual risks decline through maturation, acquired robustness, or developmental timing, and our analysis cannot distinguish these mechanisms from selection. Second, the construction retains the Siler model's assumption that its three risks act independently, and the cause-deleted decomposition inherits this assumption. Third, the Swedish estimates are conditional on the fitted Siler model, which overstates $e_0$ by up to 2.6 years before 1940.

The immaturity component of the Siler model is a Poisson-distributed addition to the Makeham term. Results known for the Gompertz--Makeham model therefore extend to the whole life course, and the life expectancy lost to immaturity mortality becomes a simple quantity with an explicit bound.

\subsection*{Funding}
This work was funded by the European Union (ERC, Born Once -- Die Once, Grant agreement ID 101043983). Views and opinions expressed are those of the author only and do not necessarily reflect those of the European Union or the European Research Council Executive Agency. Neither the European Union nor the granting authority can be held responsible for them.

\subsection*{Data and code availability}
Female death counts, exposures, and life-table values were obtained from the Human Mortality Database. The underlying data are available from \url{https://www.mortality.org/} under its access conditions. The materials required to reproduce the analyses, tables, and figures are openly available at \url{https://github.com/scpatricio/siler-poisson-mixture/}.

\subsection*{Author contributions}
Silvio C. Patricio: Conceptualization, Methodology, Software, Validation, Formal analysis, Investigation, Writing -- original draft, Writing -- review and editing, and Visualization.

{\footnotesize
\bibliography{references}
}

%%%%%%%%%%%%%%%% START OF SUPPLEMENT %%%%%%%%%%%%%%%
\newpage

% Figures, tables, equations and pages in the supplement are numbered S1, S2 etc.

\renewcommand{\thefigure}{S\arabic{figure}}
\renewcommand{\thetable}{S\arabic{table}}
\renewcommand{\theequation}{S\arabic{equation}}
\renewcommand{\thepage}{S\arabic{page}}
\setcounter{figure}{0}
\setcounter{table}{0}
\setcounter{equation}{0}
\setcounter{page}{1} % not 0 as \newpage already started a supplementary page

\appendix
\setcounter{proposition}{0}
\renewcommand{\theproposition}{S\arabic{proposition}}
\renewcommand{\theHproposition}{S\arabic{proposition}}
\setcounter{corollary}{0}
\renewcommand{\thecorollary}{S\arabic{corollary}}
\renewcommand{\theHcorollary}{S\arabic{corollary}}

\section{Extensions beyond the Siler specification}\label{app:extensions}

The clock argument of Section~\ref{sec:mixture} uses neither the exponential form of the immaturity hazard nor the Gompertz form of the senescence hazard. We therefore state it for a general immaturity hazard $\mu_1$ and a general adult hazard $h$ in place of $\mu_3$, so that the total hazard is $\mu(t)=\mu_1(t)+a_2+h(t)$. These results generalize the representation and are not proposals for new mortality models. Both $\mu_1$ and $h$ are nonnegative and locally integrable. The immaturity hazard has finite and positive cumulative hazard $A_0=\int_0^\infty\mu_1(u)\,\dd u$, so $A_x=\int_x^\infty\mu_1(u)\,\dd u$ is finite for every $x\geq0$. If $A_x=0$, as happens at ages beyond which $\mu_1$ vanishes, survivors at age $x$ carry no clocks, $K_x=0$ almost surely, and the clock terms below equal one. For a constant $c\geq0$, we write
\begin{equation}
e_{x,h}(c)=\int_0^\infty
\exp\left\{-ct-\int_x^{x+t}h(u)\,\dd u\right\}\,\dd t
\label{eq:adult-rle}
\end{equation}
for remaining life expectancy under the adult hazard $h$ and the additive constant hazard $c$. We assume $e_{x,h}(a_2)<\infty$, which holds, for instance, when $a_2>0$ or when $h$ is bounded below by a positive constant. Because $\mu_1\geq0$, the remaining life expectancy $e_x$ is then finite as well.

\subsection{Clocks with age-dependent ring times}\label{app:general-clocks}

Any immaturity hazard with finite cumulative hazard admits a clock representation if the clocks are allowed to ring at age-dependent rates. Let $N_0\sim\operatorname{Poisson}(A_0)$ and, conditional on $N_0$, let the clocks ring at independent times $C_1,\ldots,C_{N_0}$ with common density $\mu_1(t)/A_0$, independently of the background and adult failure times. The first ring $C=\min_jC_j$, with $C=\infty$ when $N_0=0$, has survival function
\[
\Pr(C>t)=\exp\left\{-A_0\left(1-\frac{A_t}{A_0}\right)\right\}
=\exp\left\{-\int_0^t\mu_1(u)\,\dd u\right\}.
\]
Each clock has not rung by age $x$ with probability $A_x/A_0$, so the Bayes argument of Section~\ref{sec:mixture} shows that a survivor at age $x$ carries $K_x\sim\operatorname{Poisson}(A_x)$ clocks. When $A_x>0$, their residual ring times are independent, with survival function $A_{x+t}/A_x$ and, wherever $A_{x+t}>0$, hazard
\[
\rho_x(t)=\frac{\mu_1(x+t)}{A_{x+t}}.
\]

\begin{proposition}[Clock representation]\label{prop:clock-general}
For every $x,t\geq0$,
\begin{equation}
{}_tp_x=\E\!\left[\left(\frac{A_{x+t}}{A_x}\right)^{K_x}
\exp\left\{-a_2t-\int_x^{x+t}h(u)\,\dd u\right\}\right],
\qquad K_x\sim\operatorname{Poisson}(A_x),
\label{eq:clock-general}
\end{equation}
and
\begin{equation}
e_x=\e^{-A_x}\sum_{n=0}^\infty\frac{(A_x)^n}{n!}
\int_0^\infty\left(\frac{A_{x+t}}{A_x}\right)^n
\exp\left\{-a_2t-\int_x^{x+t}h(u)\,\dd u\right\}\,\dd t.
\label{eq:clock-general-series}
\end{equation}
\end{proposition}

\begin{proof}
Conditional on $K_x=n$, the residual ring times are independent with survival function $A_{x+t}/A_x$, so none of them rings in the next $t$ years with probability $(A_{x+t}/A_x)^n$. Multiplying this factor by the background and adult survival functions and averaging over $K_x$ gives Equation~\eqref{eq:clock-general}. Integrating over $t$ and applying Tonelli's theorem gives Equation~\eqref{eq:clock-general-series}.
\end{proof}

Component $n$ in Equation~\eqref{eq:clock-general-series} is a remaining life expectancy under the total hazard $a_2+h(x+t)+n\rho_x(t)$. The clock hazard $\rho_x$ generally varies with time since age $x$, so the components are not obtained by adding a constant to the background hazard.

\subsection{Constant-rate clocks}\label{app:constant-rate}

The clocks of Proposition~\ref{prop:clock-general} may have hazards that change with age. The selection reading of Section~\ref{sec:mixture} needs more: each clock rings at a constant rate, although rates may differ between clocks. A clock whose rate is a positive random variable $R$ with distribution function $F$ rings at a time with density $\E(R\e^{-tR})=\int_0^\infty z\e^{-tz}\,\dd F(z)$, a mixture of exponential densities. We call $\mu_1$ completely monotone if its derivatives alternate in sign, that is, $(-1)^k\mu_1^{(k)}(t)\geq0$ for every $k\geq0$ and $t>0$.

\begin{proposition}[Constant-rate clocks]\label{prop:constant-rate}
The immaturity hazard is generated by Poisson clocks whose constant rates are independent draws from a common distribution if and only if $\mu_1$ is completely monotone. In that case there is a positive random rate $R$, unique in distribution, such that
\[
\mu_1(t)=A_0\,\E(R\e^{-tR}),
\qquad
A_x=A_0\,\E(\e^{-xR}).
\]
A survivor at age $x$ carries $K_x\sim\operatorname{Poisson}(A_x)$ clocks whose rates $\Lambda_{x,1},\Lambda_{x,2},\ldots$ are independent of $K_x$ and independent and identically distributed as a random rate $\Lambda_x$ whose distribution satisfies
\[
\E\{g(\Lambda_x)\}=\frac{\E\{g(R)\e^{-xR}\}}{\E(\e^{-xR})}
\]
for every bounded measurable function $g$. With $Y_x=\sum_{j=1}^{K_x}\Lambda_{x,j}$, and the empty sum equal to zero,
\begin{equation}
{}_tp_x=\E\!\left[\exp\left\{-(a_2+Y_x)t-\int_x^{x+t}h(u)\,\dd u\right\}\right],
\qquad
e_x=\E\!\left[e_{x,h}(a_2+Y_x)\right].
\label{eq:constant-rate-mixture}
\end{equation}
\end{proposition}

\begin{proof}
For the necessity, differentiating $\E(R\e^{-tR})$ $k$ times gives $(-1)^k\E(R^{k+1}\e^{-tR})$, so every mixture of exponential densities is completely monotone. For the sufficiency, Bernstein's theorem \citep{feller1971introduction} writes a completely monotone $\mu_1$ as $\mu_1(t)=\int_0^\infty\e^{-tz}\,\dd\Psi(z)$, with $\Psi$ nondecreasing and determined by $\mu_1$. Tonelli's theorem gives $A_0=\int_0^\infty z^{-1}\,\dd\Psi(z)$, so a finite $A_0$ excludes a jump of $\Psi$ at $z=0$, and $F(z)=A_0^{-1}\int_0^z u^{-1}\,\dd\Psi(u)$ is a distribution function on $(0,\infty)$. A rate $R$ with distribution function $F$ satisfies $\mu_1(t)=A_0\E(R\e^{-tR})$, and integrating from $x$ to $\infty$ gives $A_x=A_0\E(\e^{-xR})$. A clock with rate $z$ has not rung by age $x$ with probability $\e^{-xz}$, so Bayes' rule reweights the rates of the clocks carried at age $x$ by $\e^{-xR}$, which gives the law of $\Lambda_x$. Proposition~\ref{prop:clock-general} gives $K_x\sim\operatorname{Poisson}(A_x)$. By memorylessness, conditional on $K_x$ and the rates, none of the clocks rings in the next $t$ years with probability $\e^{-tY_x}$. Multiplying by the background and adult survival functions, averaging, and integrating with Tonelli's theorem gives Equation~\eqref{eq:constant-rate-mixture}.
\end{proof}

Proposition~\ref{prop:constant-rate} restates a classical result in mortality terms. The cumulative immaturity hazard $A_0\{1-\E(\e^{-tR})\}$ is a bounded Bernstein function, the Laplace exponent of a compound Poisson process whose jumps have distribution function $F$. A Bernstein function is a nonnegative function with a completely monotone derivative, and here that derivative is $\mu_1$ \citep{schilling2026bernstein}. The reweighting by $\e^{-xR}$ is selection among clocks: fast clocks ring early, so the clocks still carried at older ages are slower on average. Proposition~\ref{prop:constant-rate} therefore sets the limit of the selection reading. A completely monotone immaturity hazard can always be read as selection among individuals carrying constant-rate clocks, whereas any other immaturity hazard requires clocks whose hazard changes with age.

In the Siler model, $R=b_1$ with probability one, so every clock has rate $b_1$ and $Y_x=b_1K_x$. Proposition~\ref{prop:constant-rate} then reduces to Proposition~\ref{prop:distribution-mixture}, with the senescence hazard $\mu_3$ replaced by a general adult hazard $h$.

The stretched-exponential immaturity hazard $\mu_1(t)=a_1\e^{-b_1t^p}$, with $a_1,b_1,p>0$, shows where the condition binds. Its proportional rate of decline, $-\dd\log\mu_1(t)/\dd t=b_1p\,t^{p-1}$, decreases with age when $0<p<1$, is constant when $p=1$, and increases with age when $p>1$. The quantity $(b_1)^{-1/p}$ is the age at which the hazard has fallen from $a_1$ to $a_1/\e$, and the remaining cumulative hazard is
\[
A_x=\frac{a_1}{p (b_1)^{1/p}}
\Gamma\!\left(\frac{1}{p},b_1x^p\right).
\]
For $0<p<1$, $\e^{-b_1t^p}=\E(\e^{-tZ_p})$ for a positive stable random variable $Z_p$ \citep{Pollard1946}, so $\mu_1$ is completely monotone, and $R$ has the distribution of $Z_p$ reweighted by $1/Z_p$. The clock rates then have a continuous distribution, and the mixing distribution of $Y_x$ retains an atom at zero. For $p>1$, the second derivative of $\e^{-b_1t^p}$ is negative for all sufficiently small positive $t$, so $\mu_1$ is not completely monotone and constant-rate clocks are unavailable. Proposition~\ref{prop:clock-general} still applies, with clocks whose hazard $\rho_x$ varies with age.

\subsection{Exponential immaturity and general adult hazards}\label{app:general-adult}

With exponential immaturity mortality, every clock has the same constant rate, and the truncation bound of Section~\ref{sec:truncation} carries over to any adult hazard.

\begin{corollary}[General adult hazard]\label{cor:general-adult}
If $\mu_1(t)=a_1\e^{-b_1t}$, then
\begin{equation}
e_x=\E\!\left[e_{x,h}(a_2+b_1K_x)\right],
\qquad K_x\sim\operatorname{Poisson}(A_x).
\label{eq:general-adult-mixture}
\end{equation}
The components $Q_{n,h}(x)=e_{x,h}(a_2+nb_1)$ are nonincreasing in $n$ and satisfy $Q_{n,h}(x)\leq1/(a_2+nb_1)$ whenever the denominator is positive, so the truncation bound in Equation~\eqref{eq:truncation-bound} holds with $Q_n(x)$ replaced by $Q_{n,h}(x)$.
\end{corollary}

\begin{proof}
Equation~\eqref{eq:general-adult-mixture} is Proposition~\ref{prop:constant-rate} with $R=b_1$ with probability one. The integrand in Equation~\eqref{eq:adult-rle} decreases in $c$ and is bounded by $\e^{-ct}$, which gives the monotonicity and the bound. The truncation argument is then that of Proposition~\ref{prop:truncation}.
\end{proof}

For the gamma--Gompertz case, set
\[
h(t)=\frac{\alpha\e^{\beta t}}
{1+\sigma^2(\alpha/\beta)(\e^{\beta t}-1)},
\qquad \alpha,\beta,\sigma^2>0,
\]
and write $e_x^{\mathrm{GGM}}(c)=e_{x,h}(c)$ for gamma--Gompertz--Makeham remaining life expectancy with Makeham term $c$. Define
\[
q=\frac{\sigma^2\alpha}{\beta},
\qquad
q_x=\frac{q\e^{\beta x}}{1+q(\e^{\beta x}-1)},
\qquad
s_c=\frac{c}{\beta}+\frac{1}{\sigma^2}.
\]
The conditional gamma--Gompertz survival function is
\[
\left[1+q_x(\e^{\beta t}-1)\right]^{-1/\sigma^2}.
\]
Substituting $y=\e^{-\beta t}$ into Equation~\eqref{eq:adult-rle} gives
\begin{align*}
e_x^{\mathrm{GGM}}(c)
&=\frac{1}{\beta}\int_0^1
y^{s_c-1}\{q_x+(1-q_x)y\}^{-1/\sigma^2}\,\dd y \\
&=\frac{(q_x)^{-1/\sigma^2}}{\beta s_c}
{}_2F_1\!\left(\frac{1}{\sigma^2},s_c;s_c+1;
1-\frac{1}{q_x}\right).
\end{align*}
This expression agrees with the closed form derived by \citet{CastellaresPatricioLemonte2024}.

\begin{corollary}[Gamma--Gompertz adult mortality]\label{cor:gamma-gompertz}
For the gamma--Gompertz adult hazard above,
\[
e_x=\E\!\left[e_x^{\mathrm{GGM}}(a_2+b_1K_x)\right],
\qquad K_x\sim\operatorname{Poisson}(A_x).
\]
The truncation bound in Equation~\eqref{eq:truncation-bound} also holds with $Q_n(x)$ replaced by $e_x^{\mathrm{GGM}}(a_2+nb_1)$.
\end{corollary}

\begin{proof}
Substitute $e_{x,h}(c)=e_x^{\mathrm{GGM}}(c)$ into Corollary~\ref{cor:general-adult}.
\end{proof}

\clearpage
\section{Numerical verification}\label{app:numerical-verification}

\begin{figure}[!htb]
\centering
\includegraphics[width=\textwidth]{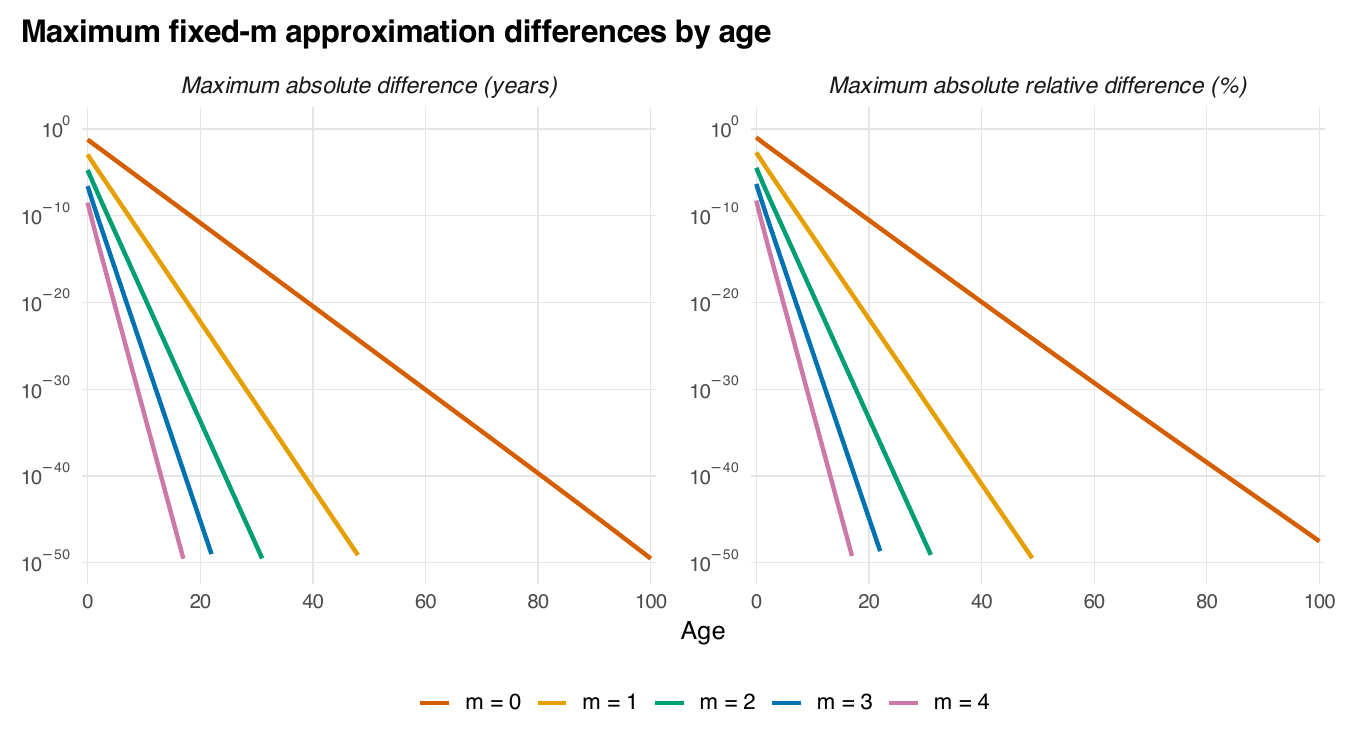}
\caption{Largest truncation error of the Siler series across all fitted years, Swedish females, 1900--2020: $|e_x^{(m)}-e_x|$ in years (left) and as a percentage of $e_x$ (right), $m=0,\ldots,4$. The errors are evaluated by summing the nonnegative mixture terms beyond $m$. Logarithmic scales; values below $10^{-50}$ are omitted.}
\label{fig:maximum-approximation-difference}
\end{figure}

Table~\ref{tab:numerical-verification} compares the series with adaptive quadrature in six benchmark years at ages 0, 30, and 60. The bound in the last column covers truncation only, not the error of the numerical routines. Figure~\ref{fig:maximum-approximation-difference} gives, at each age, the largest truncation error over all fitted years from 1900 to 2020.

\begingroup
\captionsetup{type=table}
\renewenvironment{table*}[1][]{\par}{\par}
\begin{minipage}{\textwidth}
\setlength{\tabcolsep}{3.5pt}%
\begin{table*}[!t]
\centering
\small
\caption{Numerical verification of the exact Siler series. The value column reports the series to four decimal places; adaptive quadrature gives the same displayed values. The absolute difference compares the series, stopped at an absolute truncation tolerance of $10^{-12}$ years, with adaptive quadrature. The final column is the analytic bound on the omitted series tail. Values below $10^{-16}$ are printed as $<10^{-16}$. Differences can exceed the bound because they also include the numerical error of the special-function routine and of the quadrature, which the bound does not cover.}
\label{tab:numerical-verification}
\begin{tabular}{rrrrr}
\toprule
Year & Age & Value & Absolute difference & Analytic bound \\
\midrule
1920 & 0 & 61.9301 & $2.70\times 10^{-13}$ & $2.64\times 10^{-13}$ \\
1920 & 30 & 40.8116 & $2.13\times 10^{-14}$ & $<10^{-16}$ \\
1920 & 60 & 17.0874 & $1.42\times 10^{-14}$ & $<10^{-16}$ \\
1940 & 0 & 69.0828 & $<10^{-16}$ & $1.59\times 10^{-13}$ \\
1940 & 30 & 42.4986 & $4.97\times 10^{-14}$ & $<10^{-16}$ \\
1940 & 60 & 16.7587 & $9.95\times 10^{-14}$ & $<10^{-16}$ \\
1960 & 0 & 75.1118 & $9.95\times 10^{-14}$ & $3.09\times 10^{-15}$ \\
1960 & 30 & 46.2970 & $2.42\times 10^{-13}$ & $<10^{-16}$ \\
1960 & 60 & 18.8274 & $5.79\times 10^{-13}$ & $<10^{-16}$ \\
1980 & 0 & 78.7106 & $7.11\times 10^{-14}$ & $4.57\times 10^{-14}$ \\
1980 & 30 & 49.4873 & $8.74\times 10^{-13}$ & $<10^{-16}$ \\
1980 & 60 & 21.5061 & $8.88\times 10^{-14}$ & $<10^{-16}$ \\
2000 & 0 & 81.7100 & $8.38\times 10^{-13}$ & $1.22\times 10^{-14}$ \\
2000 & 30 & 52.1381 & $8.88\times 10^{-13}$ & $<10^{-16}$ \\
2000 & 60 & 23.5945 & $1.08\times 10^{-12}$ & $<10^{-16}$ \\
2020 & 0 & 83.9384 & $7.11\times 10^{-13}$ & $1.92\times 10^{-16}$ \\
2020 & 30 & 54.3252 & $4.55\times 10^{-13}$ & $<10^{-16}$ \\
2020 & 60 & 25.3299 & $2.91\times 10^{-13}$ & $<10^{-16}$ \\
\bottomrule
\end{tabular}
\end{table*}

\end{minipage}
\endgroup

\end{document}